\documentclass[aps,superscriptaddress,floatfix,amsmath,amssymb]{revtex4-2}
\usepackage{inputenc}
\usepackage[a-1b]{pdfx}
\usepackage{amsmath}
\usepackage{amssymb}
\usepackage{amsthm}
\usepackage{bm}
\usepackage{graphicx}
\usepackage{subfig}
\usepackage{hyperref}
\usepackage[capitalise]{cleveref}

\newcommand\norm[1]{\left\lVert#1\right\Vert}

\begin{document}

\title{Consensus in Complex Networks with Noisy Agents and Peer Pressure}

\author{Christopher Griffin}\email{griffinch@psu.edu}
\affiliation{Applied Research Laboratory, Penn State University, University Park PA 16802} 
\author{Anna Squicciarini}\email{acs20@psu.edu}
\author{Feiran Jia}\email{fzj5059@psu.edu}
\affiliation{College of Information Science and Technology, Penn State University, University Park PA 16802}

\date{\today}

\begin{abstract}
In this paper we study a discrete time consensus model on a connected graph with  monotonically increasing peer-pressure and noise perturbed outputs masking a hidden state. We assume that each agent maintains a constant hidden state and a presents a dynamic output that is perturbed by random noise drawn from a mean-zero distribution. We show consensus is ensured in the limit as time goes to infinity under certain assumptions on the increasing peer-pressure term and also show that the hidden state cannot be exactly recovered even when model dynamics and outputs are known. The exact nature of the distribution is computed for a simple two vertex graph and results found are shown to generalize (empirically) to more complex graph structures.
\end{abstract}

\maketitle

\section{Introduction}

Many physical, biological and social systems exhibit consensus in which an internal variable is driven through interactions to a common value (see \cite{MT14} for a survey). Social consensus has been studied in  \cite{DeGroot74,Krause00,Centola15}. Swarming and flocking has been studied in \cite{TT98,CS07,EK01,L08,LX10, DM11}. Other collective motion is studied in \cite{MT14,VZ12,OFM07}. Particularly large area of research are in opinion dynamics \cite{DeGroot74,Krause00,HK02,BN05,WDA05,Toscani06,Weisb06,Lorenz07,BHT09,CFL09,KR11,DMPW12,CFT12,JM14,SZAS21,GG21} and self-organized behavior or flocking \cite{CS07,HT08,HL09,CFRT10,MOA10,Hask13,EHS16,VZ12,MT14}. Social consensus within networks has been considered in \cite{DY00,MSC01,OM04,J08,H10,DdGL10,BHT13,AALM21,WSX21,LLL22,HGHY22,HZML22,PSW22}, while similar work for natural systems evolving with network-based communication is studied in \cite{BCCC08,HH08,CCGP10,K21}. Voter models \cite{CFL09,CS73,HL75,SES05} have also been considered extensively in the statistical physics literature as models of opinion dynamics. These models have the benefit that they are solvable in restricted cases. There has also been substantial work on control and stability in these systems \cite{BHOT05,BHT09,JK10,L08,OFM07}. We also note the references given in this section are only a small snapshot of a larger literature on consensus, flocking and swarming. 

In this paper, we consider a model in which peer-pressure is a monotonic function of time \cite{SGSR18, G21} and agent outputs are subject to noise. Each agent has an internal hidden state that remains constant and a time-varying exposed state that is subject to noise as it is exchanged with its peers within a network. We assume agents prefer to release a perturbed rather than their true (hidden) state $x_0^i$ for (i) agents do not know their true state and their preferred output is drawn from a distribution centered around their true (hidden) state or (ii) agents have a privacy preference and prefer not to reveal their true state \cite{lou2017cost,10.14778/1687627.1687719}. In the latter case, the computer science literature has studied privacy extensively with some attempts to reconcile differential privacy with consensus \cite{10.1145/2381966.2381978} ultimately leading to a proof that any differentially private algorithm cannot achieve network consensus \cite{nozari2015differentially}.

We contrast these prior results by showing that as long as underlying communications network is connected and the peer-pressure is monotonic but does not increase too quickly (as given in \cite{G21}), then even with added noise consensus to an average of all hidden states is guaranteed. We also show that it is impossible to precisely recover the initial hidden states of the system, formally proving this in the case of a two vertex network and illustrating similar results empirically on a larger scale-free network.  

The remainder of this paper is organized as follows: In \cref{sec:Model} we present the model and prove that consensus to a common value is guaranteed. We analyze the problem of recovering the hidden states in \cref{sec:Hidden}. Empirical results are provided in \cref{sec:Experiment}. Results and conclusions are given in \cref{sec:Conclusions}

\section{Model}\label{sec:Model}
Let $G = (V,E)$ be a connected simple graph (no multi-edges or self-loops) with vertex set $V = \{1,\dots,n\}$ and edge set $E$. We assume each vertex represents an agent with a constant hidden state  $x^i_0 \in \mathbb{R}$ and a dynamic shared state $y^i_t$ that changes in time. Assume a symmetric edge weight function $w:E\to\mathbb{R}$ with edge $\{i,j\}$ having weight $w_{ij}$. 

At time $t$, a random value $\xi^i_t \sim \mathcal{D}$ is chosen from a distribution $\mathcal{D}^i$. For simplicity we assume $\mathbb{E}\left(\xi^i\right) = x_0^i$. Has time-varying Hamiltonian:
\begin{equation}
\mathcal{H}^i = \frac{s^i}{2}\left(\xi^i_t - y_t^i\right)^2 + \sum_{j \in N(i)}\frac{\rho_t}{2}w_{ij}\left(y_{t-1}^j - y_t^i\right)^2,
\label{eqn:Obj}
\end{equation}
which encodes the social energy of the agent experiences caused by (i) a preference to expose the random data $\xi^i$ and (ii) social pressure to agree with neighbors' previously exposed states. Here $\rho_t$ is the time varying weight placed on the term in the Hamiltonian corresponding to the social pressure. The value $s^i$ is the constant weight placed on the term in the Hamiltonian corresponding to preference for releasing random information. 

When each agent simultaneously minimizes \cref{eqn:Obj}, the resulting update rule is given by:
\begin{equation}
y^i_{t} = \frac{s^i\xi_{t}^{i} + \rho_t \sum_{j \in N(i)} w_{ij} y^j_{t-1}}{s^i + \rho_t\sum_{j \in N(i)} w_{ij}},
\label{eqn:DerivedUpdate}
\end{equation}
which is similar to the update rules in \cite{SGSR18,G21,T19,T15,Bindel2015248} using a deterministic model.

Define:
\begin{displaymath}
u^{i}_t = \frac{s^i\xi_{t}^{i} + \rho_t \sum_{j \in N(i)} w_{ij} y_{t-1}^{j}}{s^i + \rho_t\sum_{j \in N(i)} w_{ij}} - x_{0}^{i}.
\end{displaymath}
and let $\epsilon^i_t = \xi_{t}^i - \mathbf{x}_0$. By assumption, $\mathbb{E}(\epsilon^i_t) = 0$. Then we have
\begin{equation*}
u^{i_k}_t = \frac{s^i\left(x_{0}^{i} + \epsilon_t^i\right) + \rho_t \sum_{j \in N(i)} w_{ij} y_{t-1}^{j}}{s^i + \rho_t\sum_{j \in N(i)} w_{ij}} - x_{0}^{i} =
 \frac{s^i\epsilon_t^i + \rho_t \sum_{j \in N(i)} w_{ij} \left(y_{t-1}^{j} - x_{0}^{i}\right)}{s^i + \rho_t\sum_{j \in N(i)} w_{ij}}.
\end{equation*}
\cref{eqn:DerivedUpdate} can be re-written as:
\begin{equation}
y^{i}_{t} = \frac{s^i x_{0}^{i} + \rho_t \sum_{j \in N(i)} w_{ij} y_{t-1}^{j}}{s^i + \rho_t\sum_{j \in N(i)} w_{ij}} + \frac{s^i}{s^i + \rho_t\sum_{j \in N(i)} w_{ij}}\epsilon_{t}^{i}.
\label{eqn:DerivedUpdate2}
\end{equation}
From this we conclude that 
\begin{equation}
\mathbb{E}\left(y^i_t\right) = \frac{s^i x_{0}^{i} + \rho_t \sum_{j \in N(i)} w_{ij} y_{t-1}^{j}}{s^i + \rho_t\sum_{j \in N(i)} w_{ij}},
\label{eqn:Expectation}
\end{equation}
since $\mathbb{E}(\epsilon^i_t) = 0$. Following \cite{SGSR18,G21}, let $\mathbf{A}$ be the symmetric weighted adjacency matrix with $\mathbf{A}_{ij} = w_{ij}$ and let $\mathbf{D}$ be the diagonal matrix with
\begin{equation*}
\mathbf{D}_{ii} = \sum_{j} w_{ij}.
\end{equation*}
Finally, let $\mathbf{S}$ be the diagonal matrix with entries $\mathbf{S}_{ii} = s^i$. If $\mathbf{y} = \left\langle{y^1,\dots,y^n}\right\rangle$, then define
\begin{equation*}
f_t(\mathbf{y}) = \left(\mathbf{S} + \rho_t\mathbf{D}\right)^{-1}\left(\mathbf{S}\mathbf{x}_0 + \rho_t\mathbf{A}\mathbf{y}\right).
\end{equation*}
In Lemma 3 of \cite{SGSR18} it is shown that, when we fix $\rho_t$, $f_t$ is a contraction with fixed point:
\begin{equation}
\mathbf{y}^*_t = \left[\mathbf{S}+\rho_t(\mathbf{D}-\mathbf{A})\right]^{-1}\mathbf{S}\mathbf{x}_0.
\label{eqn:xkstar}
\end{equation}
To see this we solve:
\begin{equation*}
\mathbf{y} = f_t(\mathbf{y}) = \left(\mathbf{S} + \rho_t\mathbf{D}\right)^{-1}\left(\mathbf{S}\mathbf{x}_0 + \rho_t\mathbf{A}\mathbf{y}\right).
\end{equation*}
Then we have
\begin{equation*}
\left(\mathbf{S} + \rho_t\mathbf{D}\right)\mathbf{y} = \left(\mathbf{S}\mathbf{x}_0 + \rho_t\mathbf{A}\mathbf{y}\right).
\end{equation*}
Expanding both sides and subtracting $\rho_t\mathbf{A}\mathbf{y}$ gives
\begin{equation*}
\mathbf{S}\mathbf{y} + \rho_t\mathbf{D}\mathbf{y} - \rho_t\mathbf{A}\mathbf{y} = \mathbf{S}\mathbf{x}_0.
\end{equation*}
Solving for $\mathbf{y}$ gives \cref{eqn:xkstar}. The proof that $f_t$ is a contraction uses a stochastic matrix argument that is outside the scope of the paper and can be found in \cite{SGSR18}.

From Theorem IV.1 of \cite{SGSR18}, we know that if $\rho_t \to \infty$ as $t \to \infty$ and there are contraction constants $\alpha_t$ (functions of $\rho_t$) such that:
\begin{displaymath}
\norm{f_t(\mathbf{y}) - \mathbf{y}^*_t} \leq \alpha_t \norm{\mathbf{y} - \mathbf{y}_t^*} \qquad \forall \mathbf{y},
\end{displaymath}
and so that the resulting contraction constants $\alpha_t$ for $f_t$ satisfy:
\begin{displaymath}
\prod_{t=1}^\infty \alpha_t = 0,
\end{displaymath}
then for all $i$
\begin{displaymath}
\lim_{t \to \infty} y^i_t = \frac{\sum_i s_ix^i_{0}}{\sum_i s_i}.
\end{displaymath}
That is all agents come to consensus as a result of the monotonically increasing peer-pressure.

\begin{figure}[htbp]
\centering
\includegraphics[width=0.5\textwidth]{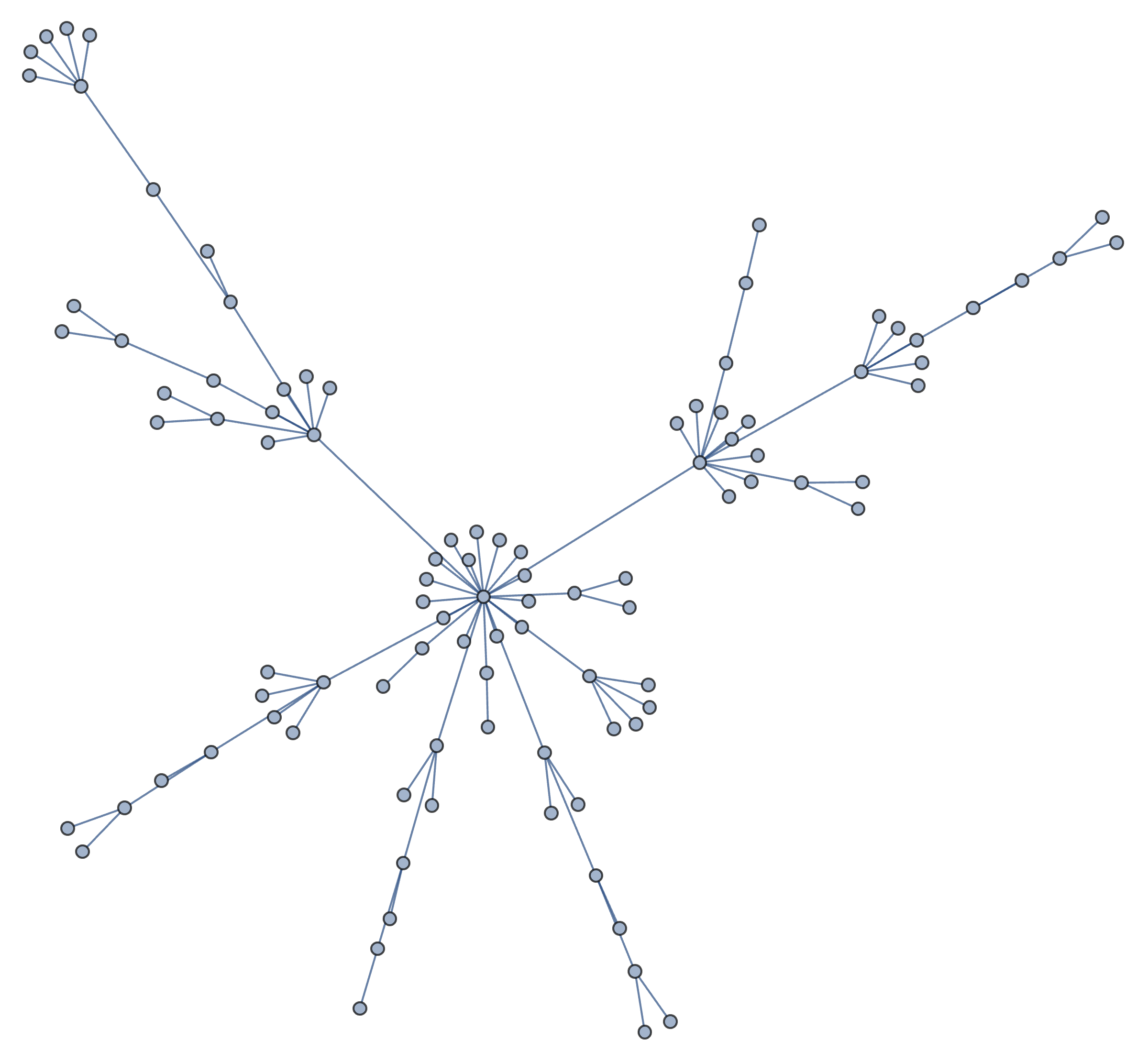}
\caption{The scale free graph used to illustrate the convergence of the dynamics with noise.}
\label{fig:ScaleFree}
\end{figure}

\begin{figure}[htbp]
\centering
\includegraphics[width=0.45\textwidth]{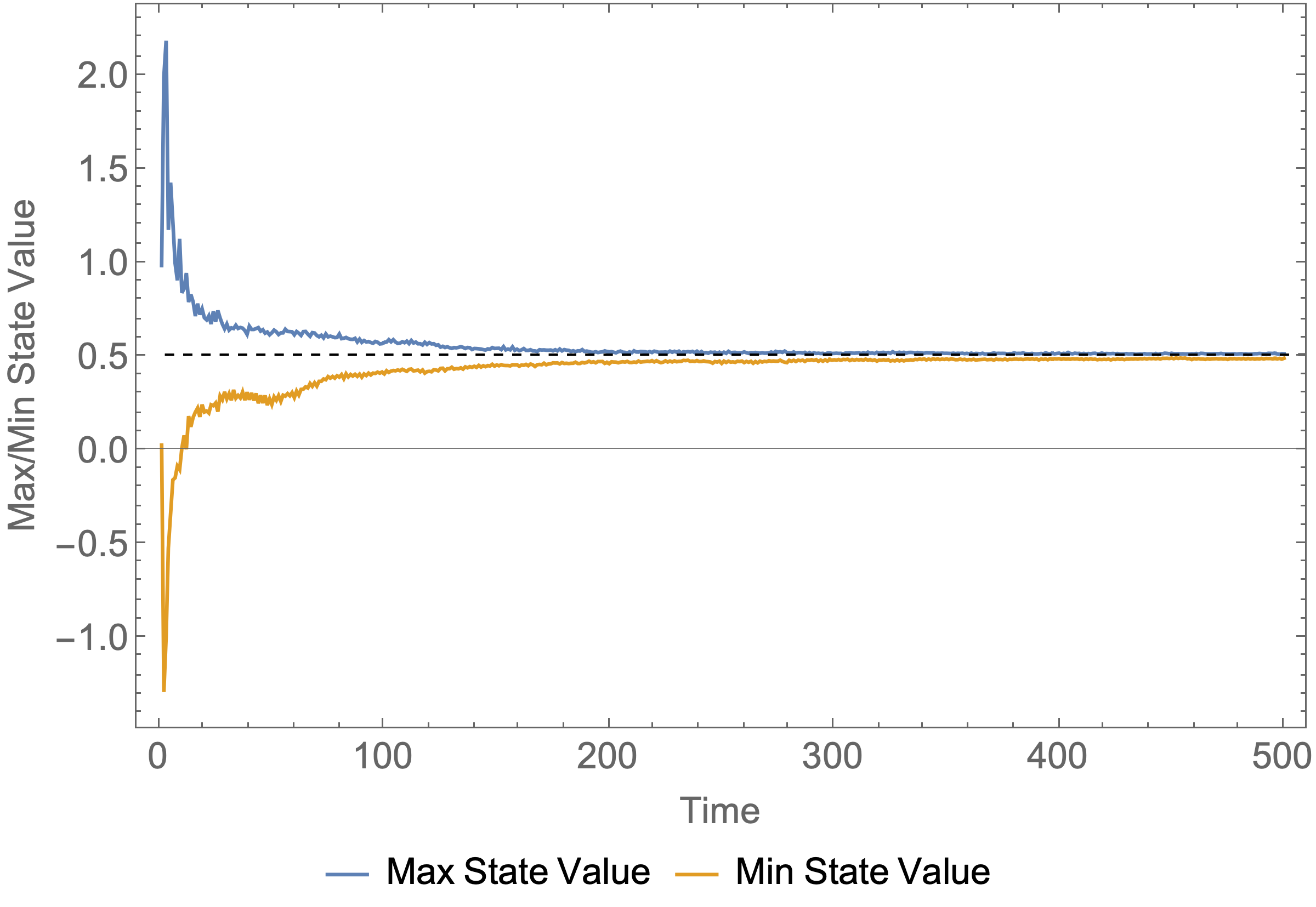} \quad
\includegraphics[width=0.45\textwidth]{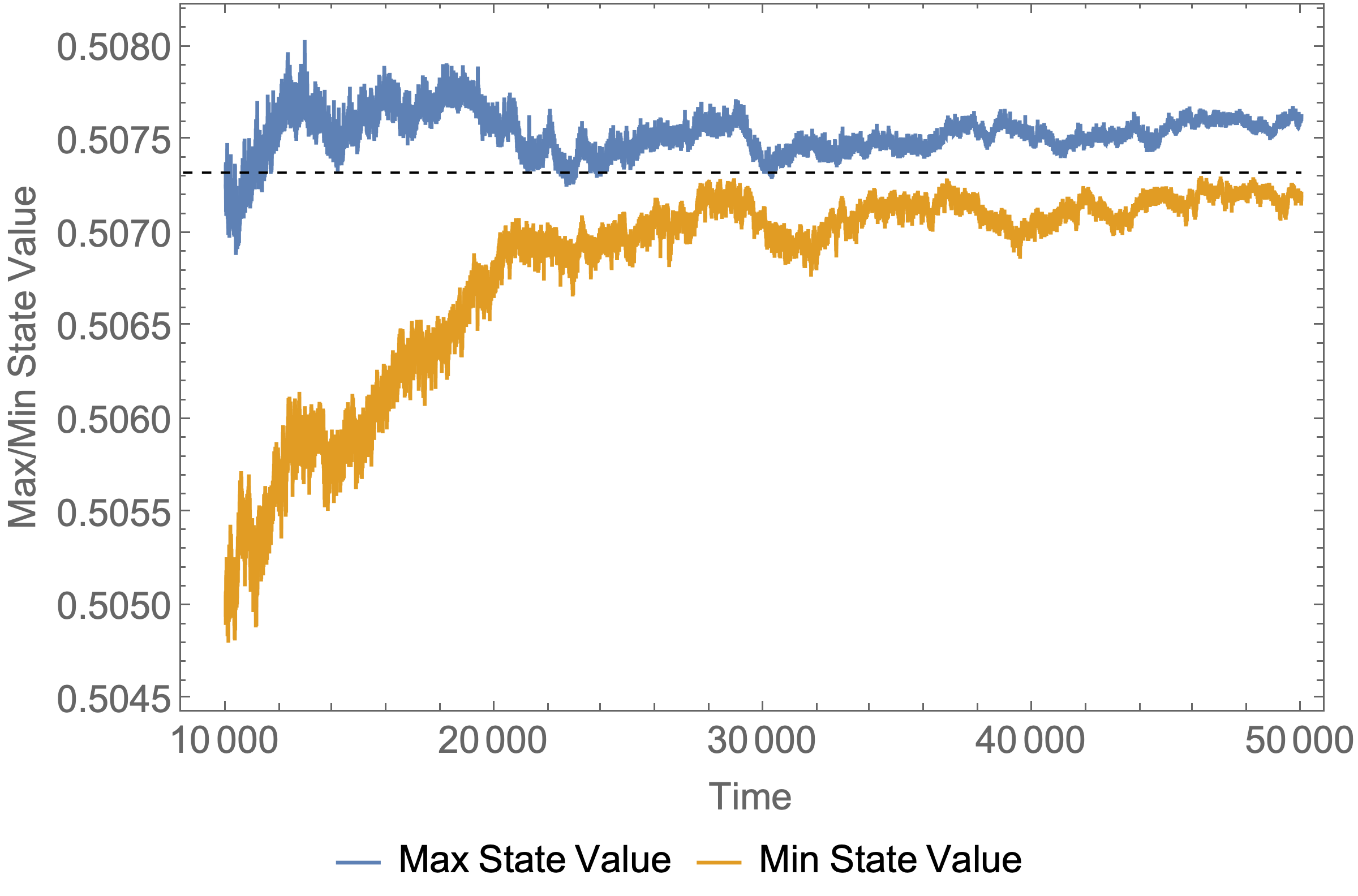} \quad
\caption{(Left) The initial time convergence of the system showing the maximum and minimum state values. (Right) Large-time convergence of the system showing maximum and minimum state values. }
\label{fig:Behavior}
\end{figure}

\begin{figure}[htbp]
\centering
\includegraphics[width=0.45\textwidth]{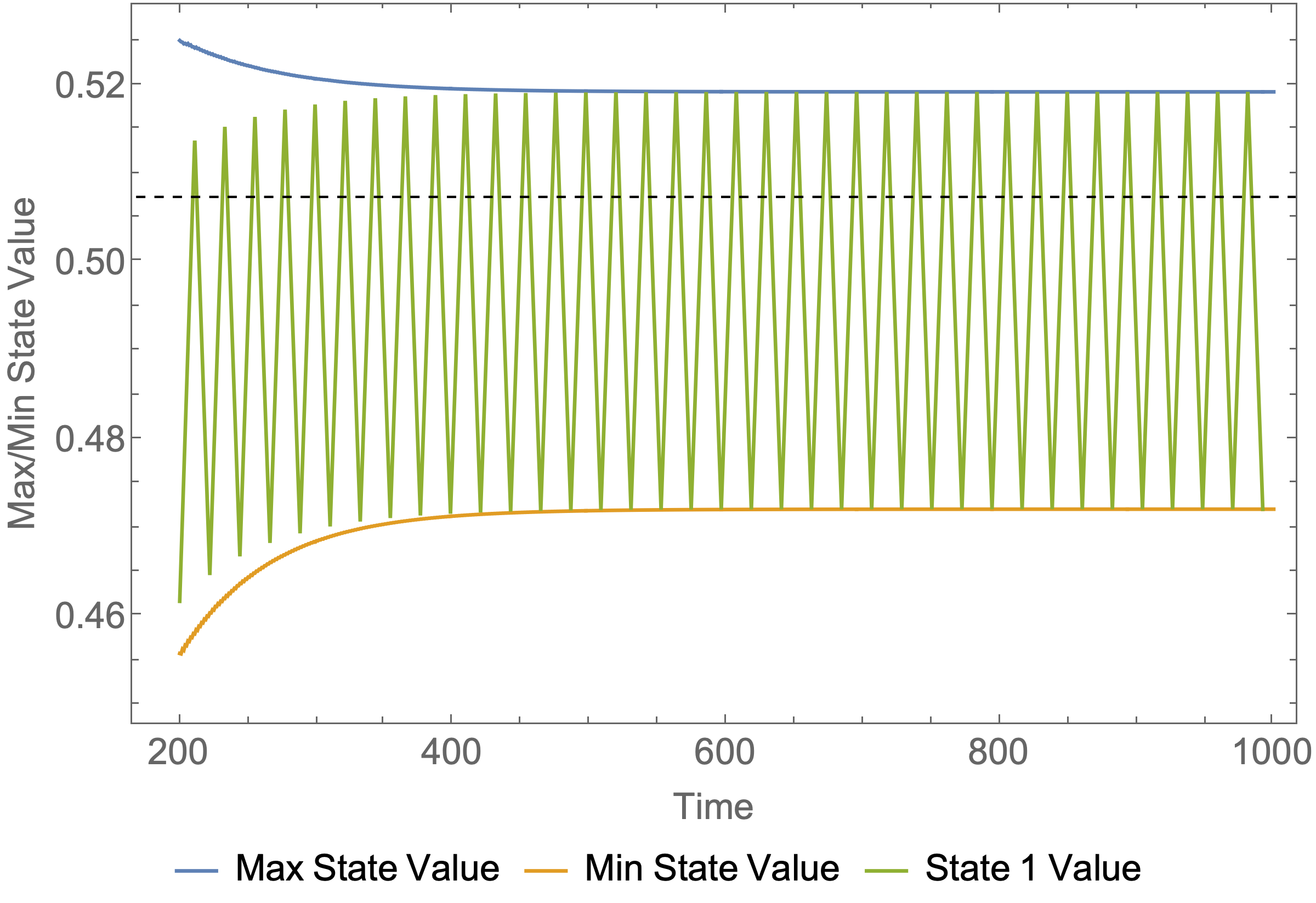}
\caption{When $\rho_t$ increases too quickly, in this case exponentially fast, convergence is not guaranteed and the system begins to oscillate.}
\label{fig:NonConvergence}
\end{figure}

Now assume $\rho_t$ grows slowly enough to ensure that $f_t$ satisfies the  conditions given above. The function $\rho_t = t$ is generally slow enough to ensure convergence. Then by \cref{eqn:Expectation}, we know that:
\begin{equation}
\lim_{t \to \infty} \mathbb{E}(y^i_t) = \frac{s^i\mathbf{x}_{0}^{i_k} + \rho_t \sum_{j \in N(i)} w_{ij} \mathbf{y}_{{t-1}}^{j_k}}{s^i + \rho_t\sum_{j \in N(i)} w_{ij}} = \frac{\sum_i s^i\mathbf{x}_0^i}{\sum_i s^i},
\end{equation}
for all $i$. Moreover, since:
\begin{equation*}
\lim_{t \to \infty}  \frac{s^i}{s^i + \rho_t\sum_{j \in N(i)} w_{ij}}\epsilon_{t}^{i} = 0,
\end{equation*}
we can conclude that that $\mathbf{y}_t$ converges not only in expectation but in value so that:
\begin{equation*}
\lim_{t \to \infty} \left\lvert y_t^i - \frac{\sum_i s^i\mathbf{x}_0^i}{\sum_i s^i}\right\vert = 0.
\end{equation*}
We illustrate this behavior using a 100 vertex randomly generated scale free graph shown in \cref{fig:ScaleFree}. We assume that $\rho_t = t$. The initial condition $\mathbf{x}_0$ was chosen at random with values in $[0,1]$. The vector $\mathbf{s}$ was chosen at random with values in $[0.5,1]$. A Mathematica notebook to perform this experiment is provided in the SI.
The resulting behavior is shown in \cref{fig:Behavior}. 
In this figure, we provide only the maximum and minimum public state values at any time. This simplifies the presentation and shows how the state converges over time without presenting 100 trajectories. We contrast this with the non-convergence when $\rho_t = 2^{\sqrt{t}}$, which grows too quickly to ensure convergence. We use the same initial condition and randomly chosen vector $\mathbf{s}$. This is illustrated in \cref{fig:NonConvergence}.
Notice that failure to converge is accompanied by system oscillation, which is illustrated by showing that the state in vertex 1 oscillates between the maximum and minimum values of all states.

\section{Recovery of the Hidden States}\label{sec:Hidden}
We analyze the problem of inferring the hidden information in $\mathbf{x}_0$ from a sequence of outputs. For simplicity, we restrict our attention to the graph $K_2$ with $w_{12} = w_{21} = 1$ and $s^1 = s^2 = 1$. We then show empirically that the theoretical results obtained for this graph hold on larger graphs. The theoretical analysis on the graph with only two vertices is (in some sense) a worse case scenario since it presents the simplest conditions under which to estimate the hidden state.

Let $(x^1_0,x^2_0)$ be the hidden state of the two players and let $(y^1_t, y^2_t)$ the current shared state. From \cref{eqn:DerivedUpdate2}, the time-varying state values can be obtained in closed form as
\begin{align}
&y^1_{t+1} = \frac{x^1_0 + y^2\rho_{t+1}}{1+\rho_{t+1}} + \frac{\epsilon_t^1}{1+\rho_{t+1}},\\
&y^2_{t+1} = \frac{x^2_0 + y^1\rho_{t+1}}{1+\rho_{t+1}} + \frac{\epsilon_t^2}{1+\rho_{t+1}}.
\end{align}


For simplicity, assume $\epsilon^i_t \sim N(0,1)$ (i.e., follows a standard normal distribution) and $\rho_t = t$, which ensures convergence \cite{G21}. Assume a stopping time $T$. If $T$ is even, then
\begin{align}
&y^1_T = \frac{\tfrac{T}{2}+1}{T+1}x^1_0 + \frac{\tfrac{T}{2}}{T+1}x^2_0 + w_T,\label{eqn:Coeff1}\\
&y^2_T = \frac{\tfrac{T}{2}}{T+1}x^1_0 + \frac{\tfrac{T}{2}+1}{T+1}x^2_0 + v_T\label{eqn:Coeff2}.
\end{align}
If $T$ is odd, then:
\begin{align}
&y^1_T = \frac{x^1_0 + x^2_0}{2} + w_T,\label{eqn:Coeff3}\\
&y^2_T = \frac{x^1_0 + x^2_0}{2} + v_T\label{eqn:Coeff4}.
\end{align}
Here $w_T$ and $v_T$ are accumulated error at $T$. Under our normality assumptions, when $T$ is even:
\begin{equation}
w_T, v_T \sim N\left(0,\sqrt{\frac{T}{2(T+1)^2}}\right).
\label{eqn:ErrorDist1}
\end{equation}
When $T$ is odd:
\begin{equation}
w_T \sim N\left(0,\sqrt{\frac{\left\lceil{T/2}\right\rceil}{(T + 1)^2}}\right) \qquad
w_T \sim N\left(0,\sqrt{\frac{\left\lfloor{T/2}\right\rfloor}{(T + 1)^2}}\right).
\label{eqn:ErrorDist2}
\end{equation}
As $T \to \infty$, $w_T, v_T \to 0$ as expected zero while:
\begin{equation*}
y^1_T, y^2_T \to \frac{x^1_0 + x^2_0}{2},
\end{equation*}
for $T$ even or odd.

We now treat this as a linear regression problem with design matrix $\mathbf{Q}$ composed of the coefficients identified for the unknown $x^1_0$ and $x^2_0$. That is we construct a vector $\mathbf{y}$ of output values with $\mathbf{y} = (y_1^1, y^2_1, \dots, y^1_T, y^2_T)$ and a the design matrix $\mathbf{Q}$ with rows given by the appropriate coefficients in \cref{eqn:Coeff1,eqn:Coeff2,eqn:Coeff3,eqn:Coeff4}. The resulting linear regression is:
\begin{displaymath}
\mathbf{y} = \begin{bmatrix} x^1_0 & x^2_0\end{bmatrix}\mathbf{Q} + \bm{\epsilon}.
\end{displaymath}
We note that the error terms in $\bm{\epsilon}$ are heteroskedastic by \cref{eqn:ErrorDist1,eqn:ErrorDist2}. However, we can use a homoskedastic assumption to construct a best case error model for the estimate of $\mathbf{x}_0 = (x^1_0,x^2_0)$. In the homoskedastic case we know that the distribution on the hidden state $\mathbf{x}_0$ is given by:
\begin{displaymath}
\hat{\mathbf{x}}_0 \sim N(\mathbf{x}_0, \bm{\Sigma}),
\end{displaymath} 
where:
\begin{equation}
\bm{\Sigma} = \hat{\sigma}\left(\mathbf{Q}\mathbf{Q}^\top\right)^{-1}.
\label{eqn:SigmaVariation}
\end{equation}
Here $\mathbf{Q}^\top$ is the matrix transpose of the design matrix $\mathbf{Q}$. The term $\hat{\sigma}$ is always a non-zero because we assume $\epsilon_t^i \sim N(0,1)$ ($i=1,2$). This would be true for any choice $\epsilon_t^i \sim N(0,\sigma_0)$, with $\sigma_0 > 0$. It now suffices to show we can compute $\left(\mathbf{Q}\mathbf{Q}^\top\right)^{-1}$. Under our assumption on $\rho_t$ and using \cref{eqn:Coeff1,eqn:Coeff2,eqn:Coeff3,eqn:Coeff4} at time $T$ we compute:
\begin{displaymath}
\mathbf{Q}\mathbf{Q}^\top = 
\begin{bmatrix} \frac{T}{2} + \sum_{t=1}^T\frac{2}{(2+4t)^2} & \frac{T}{2} - \sum_{t=1}^T\frac{2}{(2+4t)^2}\\
\frac{T}{2} - \sum_{t=1}^T\frac{2}{(2+4t)^2} &\frac{T}{2} + \sum_{t=1}^T\frac{2}{(2+4t)^2}
\end{bmatrix}.
\end{displaymath} 
We can compute the asymptotic result
\begin{displaymath}
\sum_{t=1}^\infty\frac{2}{(2+4t)^2} = \frac{\pi^2-8}{16}.
\end{displaymath}
Thus as $T\to\infty$ we have:
\begin{equation}
\bm{\Sigma} = \left(\mathbf{Q}\mathbf{Q}^\top\right)^{-1}_\infty = \left[
\begin{array}{cc}
 \frac{4}{\pi ^2-8} & \frac{4}{8- \pi ^2} \\
 \frac{4}{8- \pi ^2} & \frac{4}{\pi ^2-8} \\
\end{array}
\right].
\label{eqn:Q}
\end{equation}
This non-zero covariance matrix combined with the fact that $\hat{\sigma} > 0$ ensures that the distribution on $\mathbf{x}_0$ does not collapse to a point mass. Thus the initial data cannot be perfectly recovered. Moreover, the fact that $\bm{\Sigma} \neq \mathbf{0}$ is true even if we choose samples starting a time $t > 0$. Moreover, as the design matrix rows converge to the common value $\langle{\tfrac{1}{2},\tfrac{1}{2}}\rangle$, (see \cref{eqn:Coeff1,eqn:Coeff2,eqn:Coeff3,eqn:Coeff4}), numerical instability will make estimation of $\mathbf{x}_0$ more difficult.

\section{Empirical Results}\label{sec:Experiment}
\begin{figure}[htbp]
\centering
\includegraphics[width=0.4\textwidth]{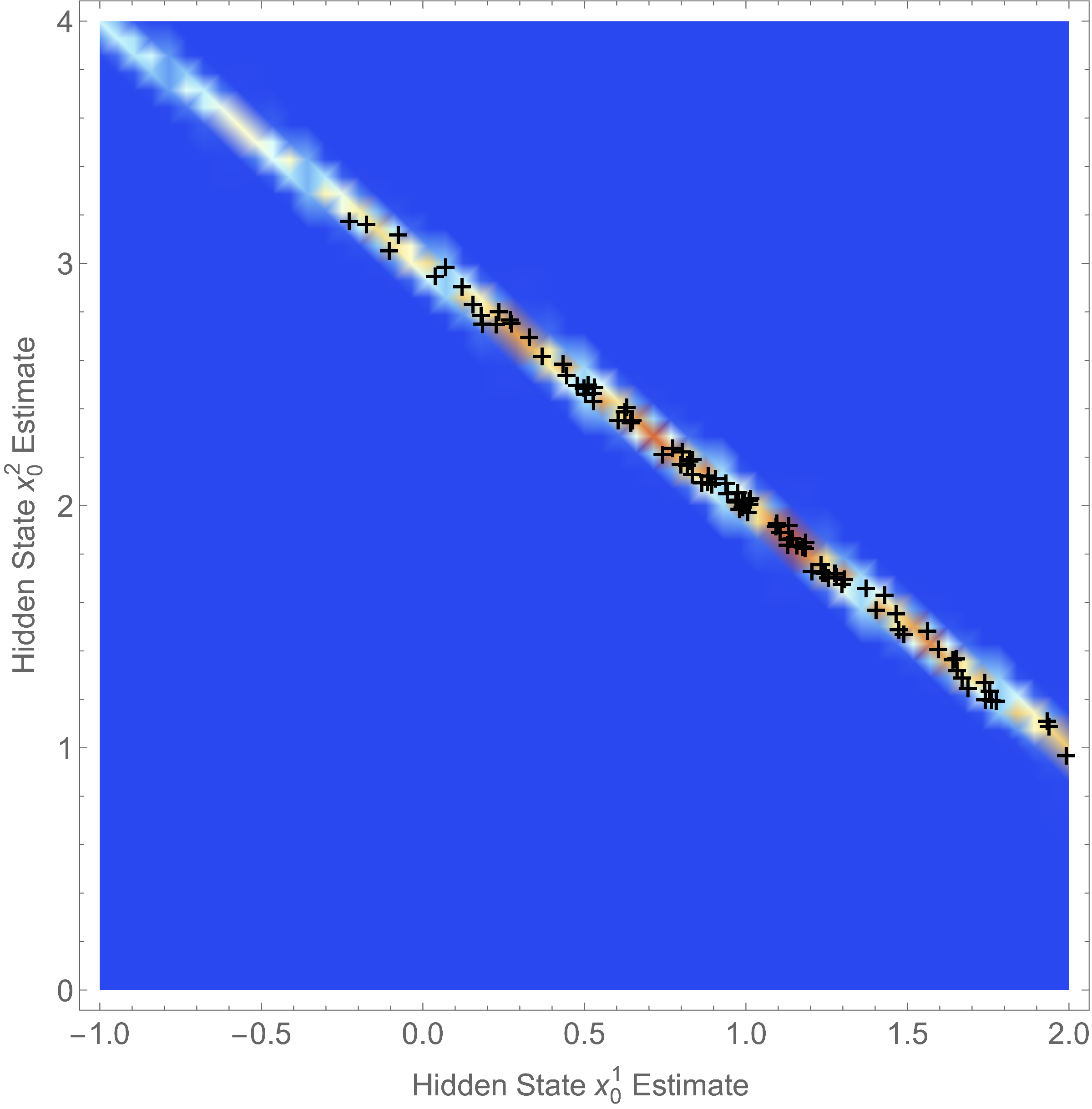}
\caption{A density plot of the asymptotic covariance matrix and the observed estimates for the two dimensional hidden state after 100 time steps.}
\label{fig:DensityPlot}
\end{figure}

In the analysis of the dynamical system on $K_2$ we noted that asymptotic result for $\bm{\Sigma}$ in \cref{eqn:Q} was developed using a homoskedastic assumption. We tested the fitting process on data generated from $K_2$ assuming that $\epsilon_t^i \sim N(0,1)$ ($i=1,2$). We used $100$ time steps and 100 replications. \cref{fig:DensityPlot} shows the 100 estimates for  $(x_0^1,x_0^2)$ along with a density plot of a normal distribution with mean $\bm{\mu}=\langle{1,2}\rangle$ (which was the initial condition used in the experiment) and $\bm{\Sigma}$ constructed from \cref{eqn:Q}.
We note that the $\bm{\Sigma}$ matrix is singular, which explains why the data are co-linear. 

\begin{figure}[htbp]
\centering
\includegraphics[width=0.445\textwidth]{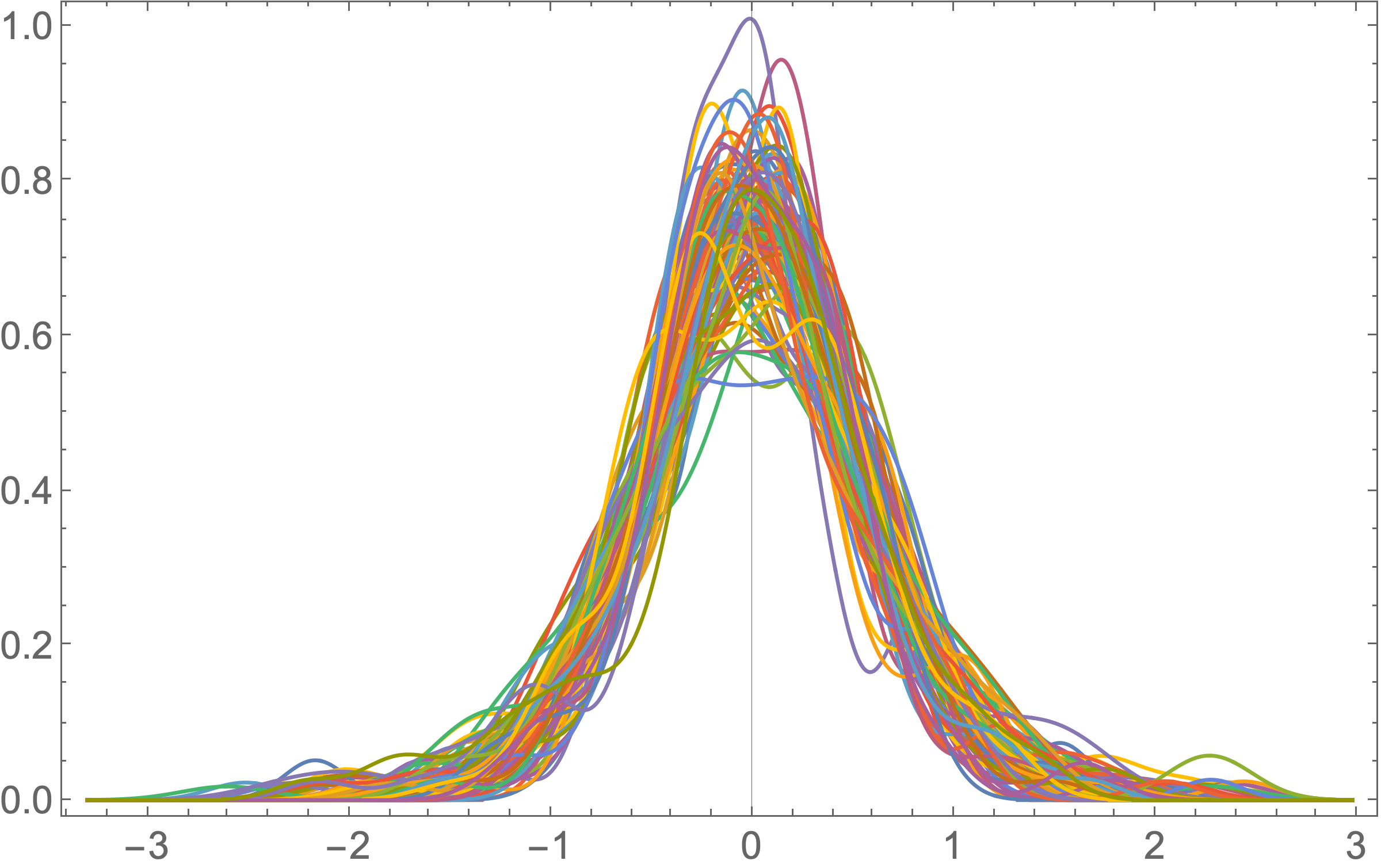}\qquad
\includegraphics[width=0.4\textwidth]{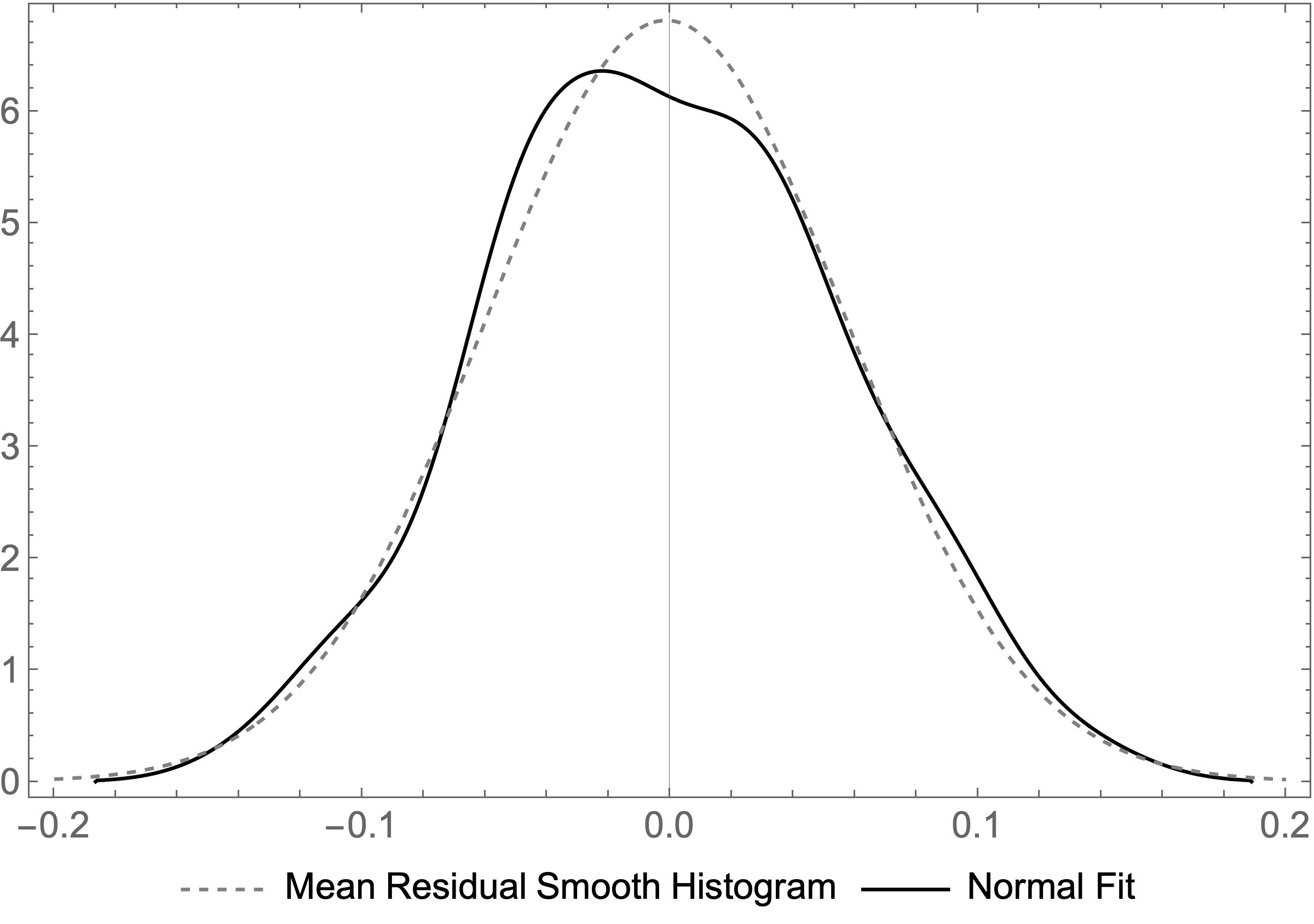} \qquad
\caption{(Left) Distributions of residuals of estimate of the hidden state from 100 agents. (Right) Mean residual distribution with normal distribution fit.}
\label{fig:ResidualAnalysis}
\end{figure}

We ran a similar experiment using a 100 vertex random graph generated according to a Barab\'{a}si-Albert distribution \cite{BA99a}. For each estimate $\hat{\mathbf{x}}_0$ of the hidden state, we constructed a residual vector $\mathbf{r}_i = \hat{\mathbf{x}}_{0_i} - \mathbf{x}_{0_i}$, (here $i = 1,\dots,100$ replications). Smoothed histograms of all residual vectors (one for each vertex) are shown in \cref{fig:ResidualAnalysis}(left). A histogram of the mean residual vector $\langle{\mathbf{r}}\rangle$ along with a normal distribution fit of the mean residual vector is shown in \cref{fig:ResidualAnalysis}(right).  



Let $\sigma_0^2$ be the variance of the noise. That is $\epsilon^i_t \sim N(0,\sigma_0)$. We tested the impact of both graph size and $\sigma_0^2$ on the variance of the residual distribution, which we denote $\sigma^2$. This distribution is illustrated in \cref{fig:ResidualAnalysis} (right). Graphs were generated with vertex size ranging from $20$ to $90$ by tens using a Barab\'{a}si-Albert (BA) distribution \cite{BA99a} with $2$ edges added in each round (rather than the usual one edge per round). The vertex range was chosen for expedience because the time to generate the fitting problem (the generalization of \cref{eqn:Coeff1,eqn:Coeff2,eqn:Coeff3,eqn:Coeff4}) is a function of graph size. We chose $\sigma_0$ from the range $[5.5,10]$ with a sample interval of $0.5$. For each pair $(|V|, \sigma_0)$ we ran 20 replications with different starting condition and measured the mean residual variance $\langle{\sigma^2}\rangle$. A second class of experiments is discussed in Appendix A on graphs generated using the standard BA distribution with $1$ edge added per round.

\begin{figure}[htbp]
\centering
\includegraphics[width=0.45\textwidth]{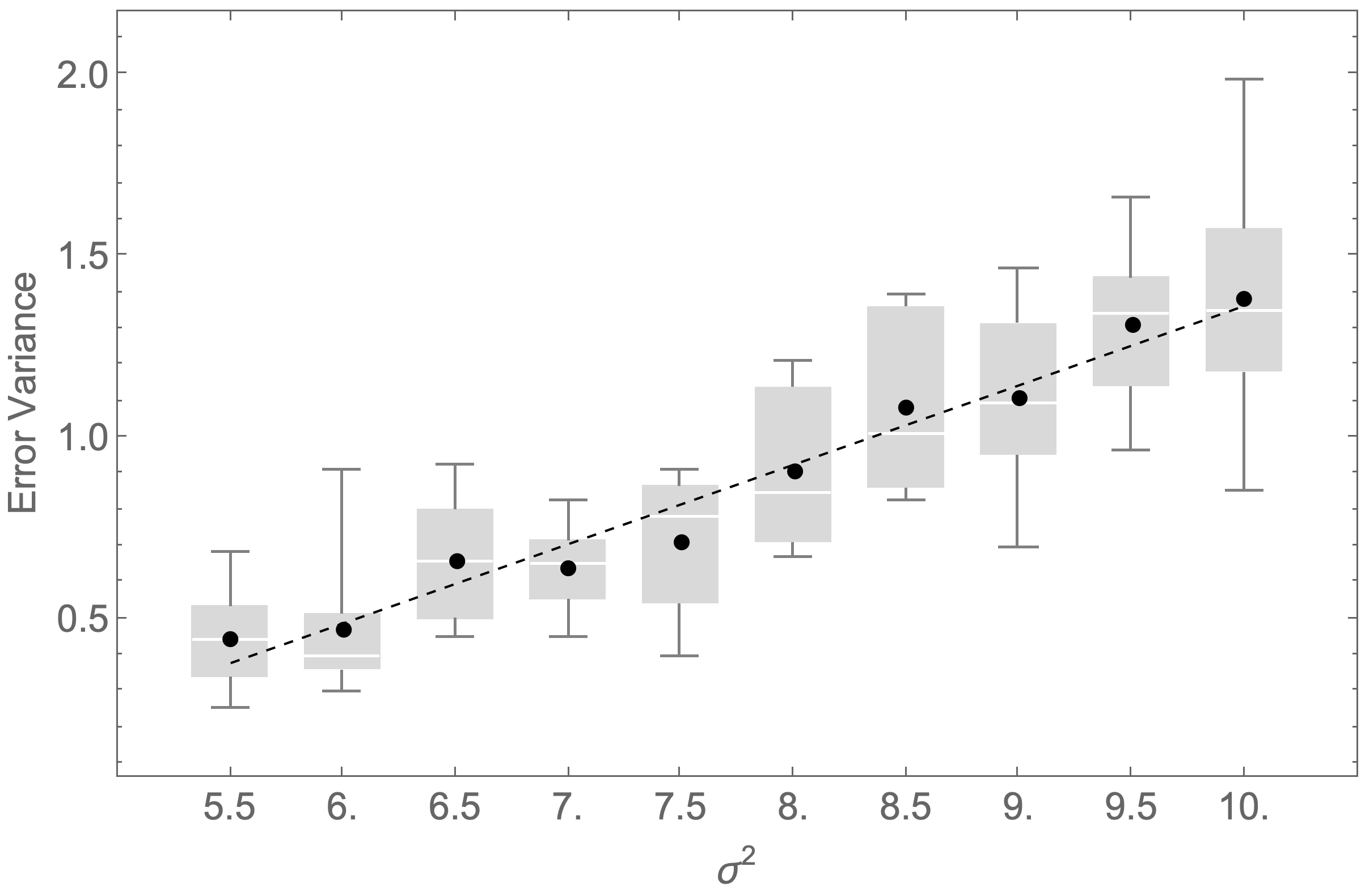}\qquad
\includegraphics[width=0.45\textwidth]{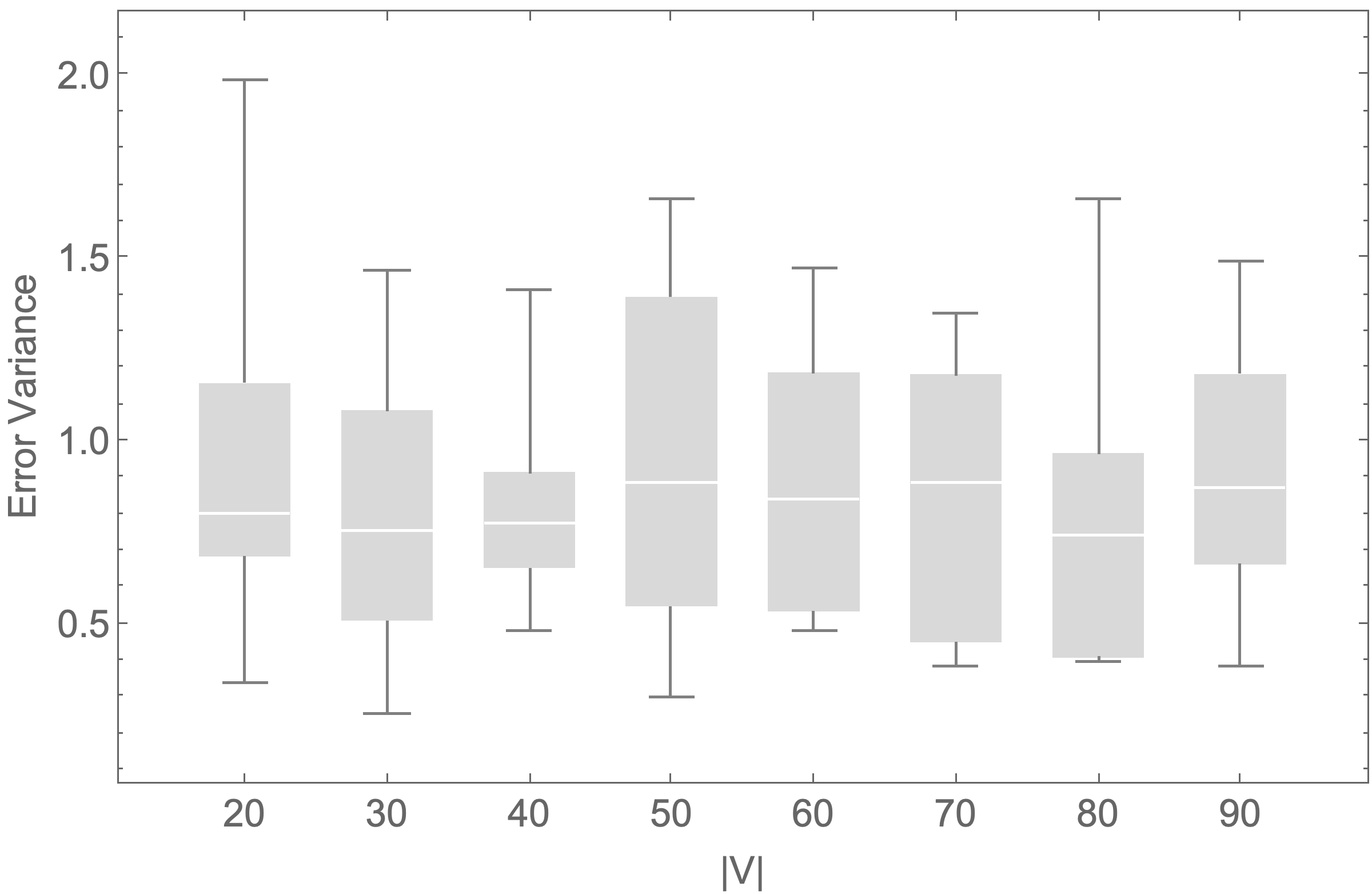}
\caption{(Left) Box-whiskers plot of the residual variances (distribution of $\sigma^2$) over 20 replications for varying noise $\sigma_0^2$ shows a linear relationship between $\hat{\sigma}^2$ and $\sigma_0^2$ as expected. (Right) Box-whiskers plot of the residual variances for varying graph sizes shows no effect from graph size.}
\label{fig:BoxWhiskers}
\end{figure}

\begin{figure}[htbp]
\centering
\includegraphics[width=0.31\textwidth]{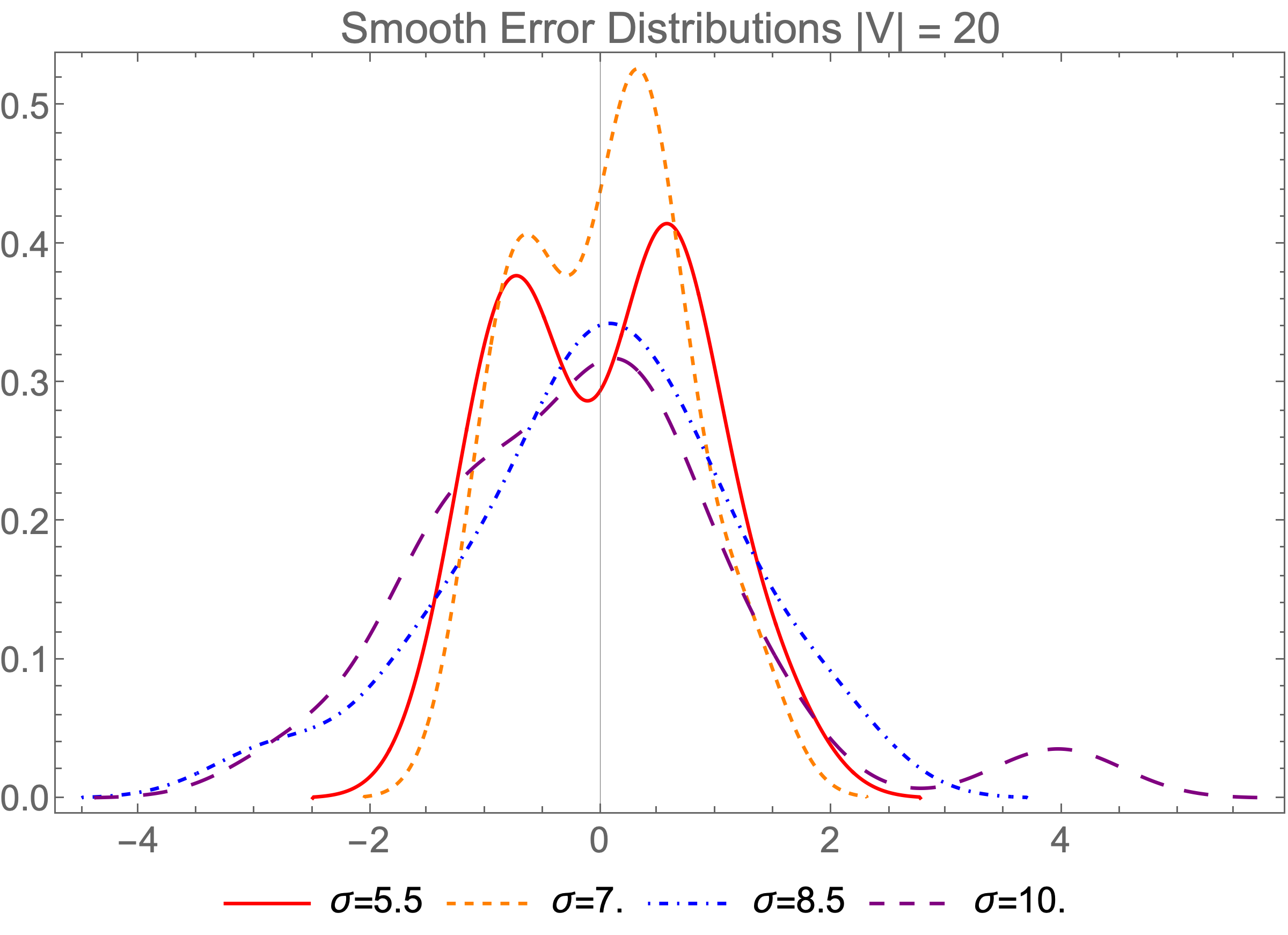}\quad
\includegraphics[width=0.31\textwidth]{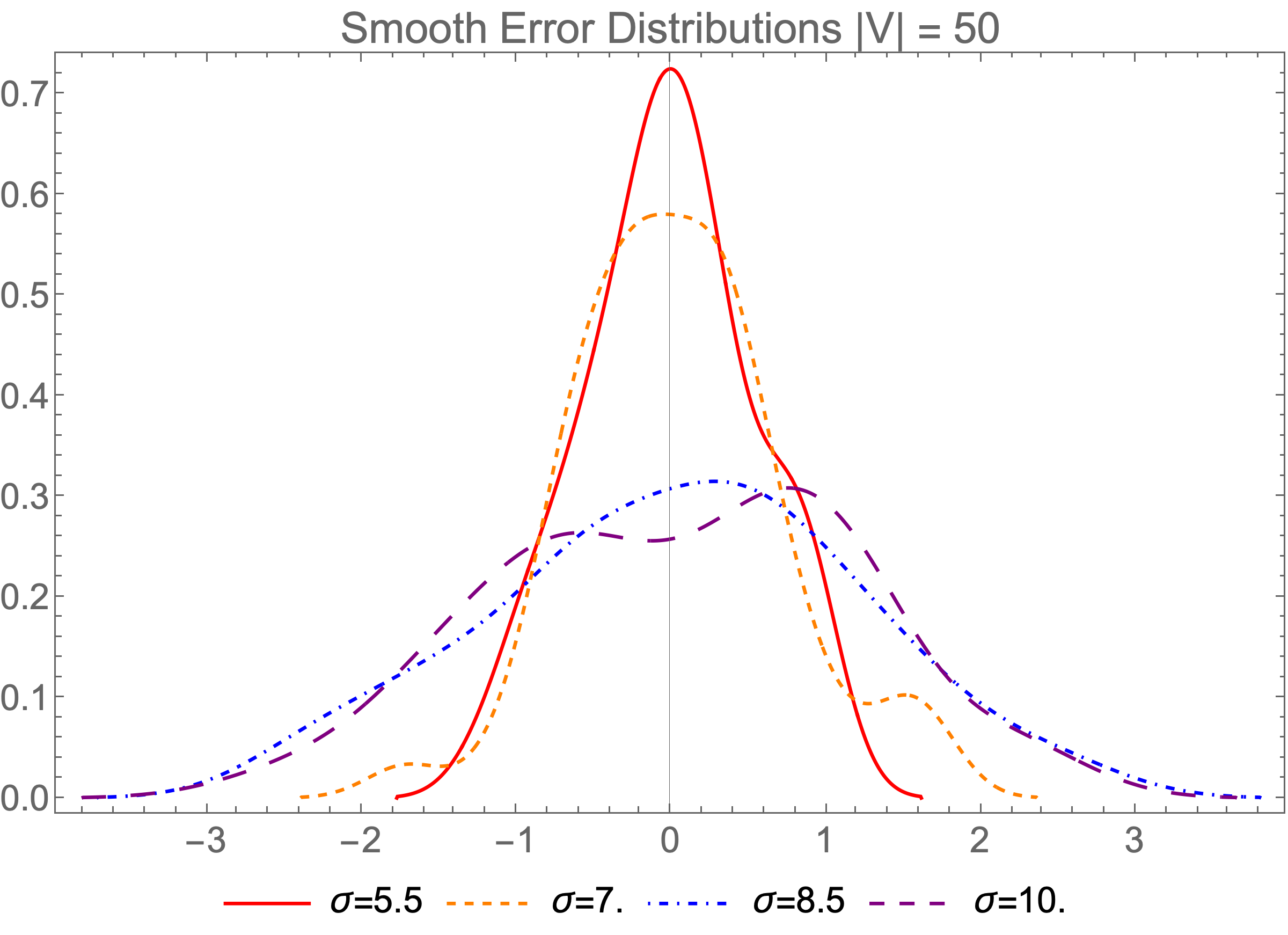}\quad
\includegraphics[width=0.31\textwidth]{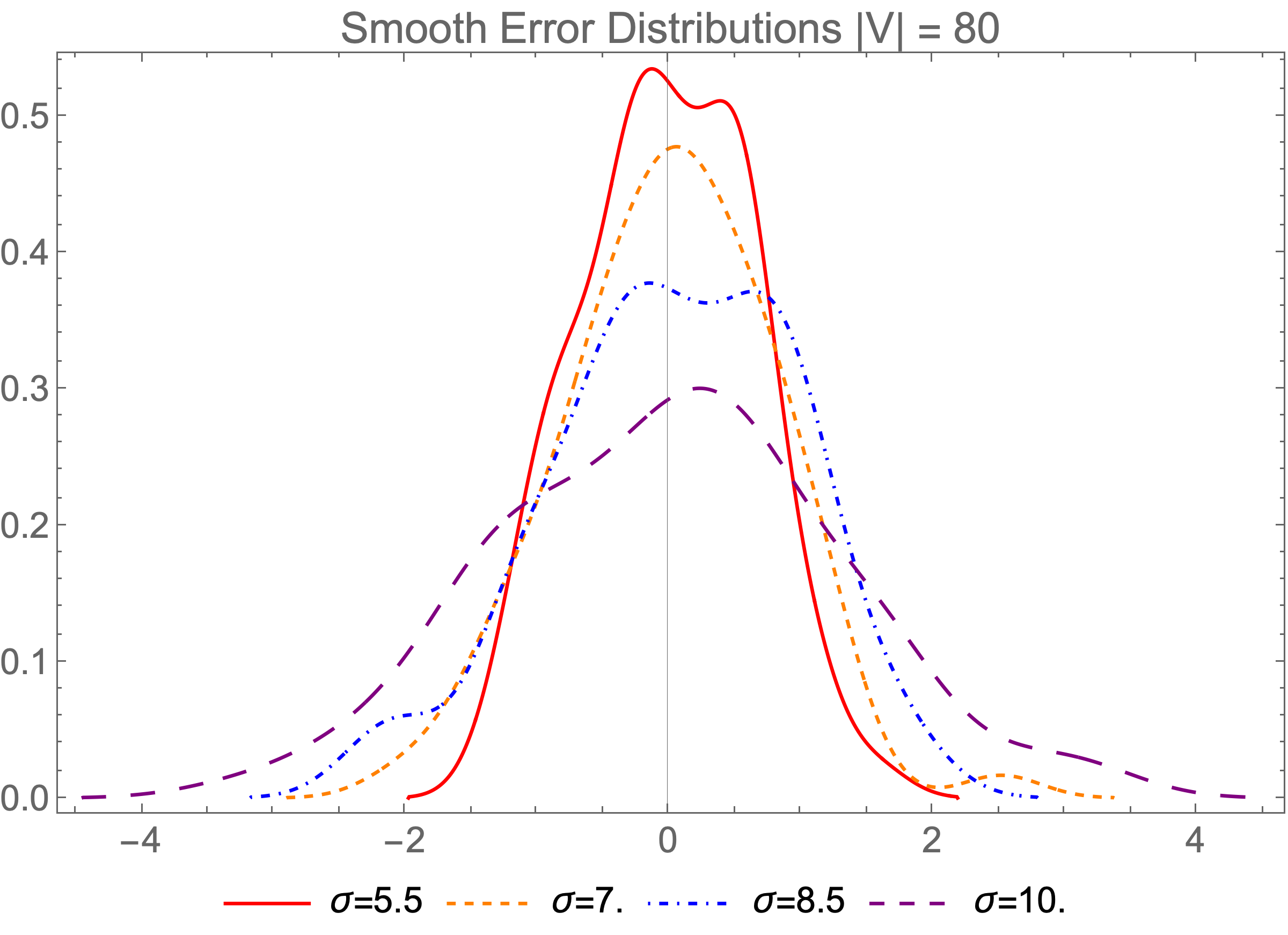}
\caption{(Left) Mean residual distributions for varying $\sigma_0^2$ when $|V| = 20$. (Middle) Mean residual distributions for varying $\sigma_0^2$ when $|V| = 50$. (Right) Mean residual distributions for varying $\sigma_0^2$ when $|V| = 80$.}
\label{fig:Distribs}
\end{figure}

When fitting the linear model
\begin{equation*}
\langle{\sigma^2}\rangle \sim \beta_0 + \beta_1|V| + \beta_2\sigma_0^2 + \epsilon,
\end{equation*}
we obtain the parameter table:
\begin{equation*}
\begin{array}{l|llll}
 \text{} & \text{Estimate} & \text{Standard Error} & \text{$t$-Statistic} & \text{$p$-Value} \\
\hline
 1 & -0.80 & 0.15 & -5.50 & 4.84\times 10^{-7} \\
 |V| & 0 & 0 & -0.41 & 0.68 \\
 \sigma_0^2  & 0.22 & 0.02 & 12.9 & 6.71\times 10^{-21}.\\
\end{array}
\end{equation*}
This suggests that the size of the graph has no impact on the residual variance, as expected. However, the mean residual variance $\langle{\sigma^2}\rangle$ grows linearly with noise variance $\sigma_0^2$ as shown in \cref{fig:BoxWhiskers}. This is expected from \cref{eqn:SigmaVariation}.
We also show the effect on mean residual distribution shape as a function of graph size $|V|$ and noise variance $\sigma_0^2$ in \cref{fig:Distribs}.
Notice the structures of the distributions are largely similar to each other (unimodal and tending toward normality) with some small-sample effects on the distributions when $|V| = 20$. This is predicted in our theoretical analysis. 

\begin{figure}[htbp]
\centering
\includegraphics[width=0.65\textwidth]{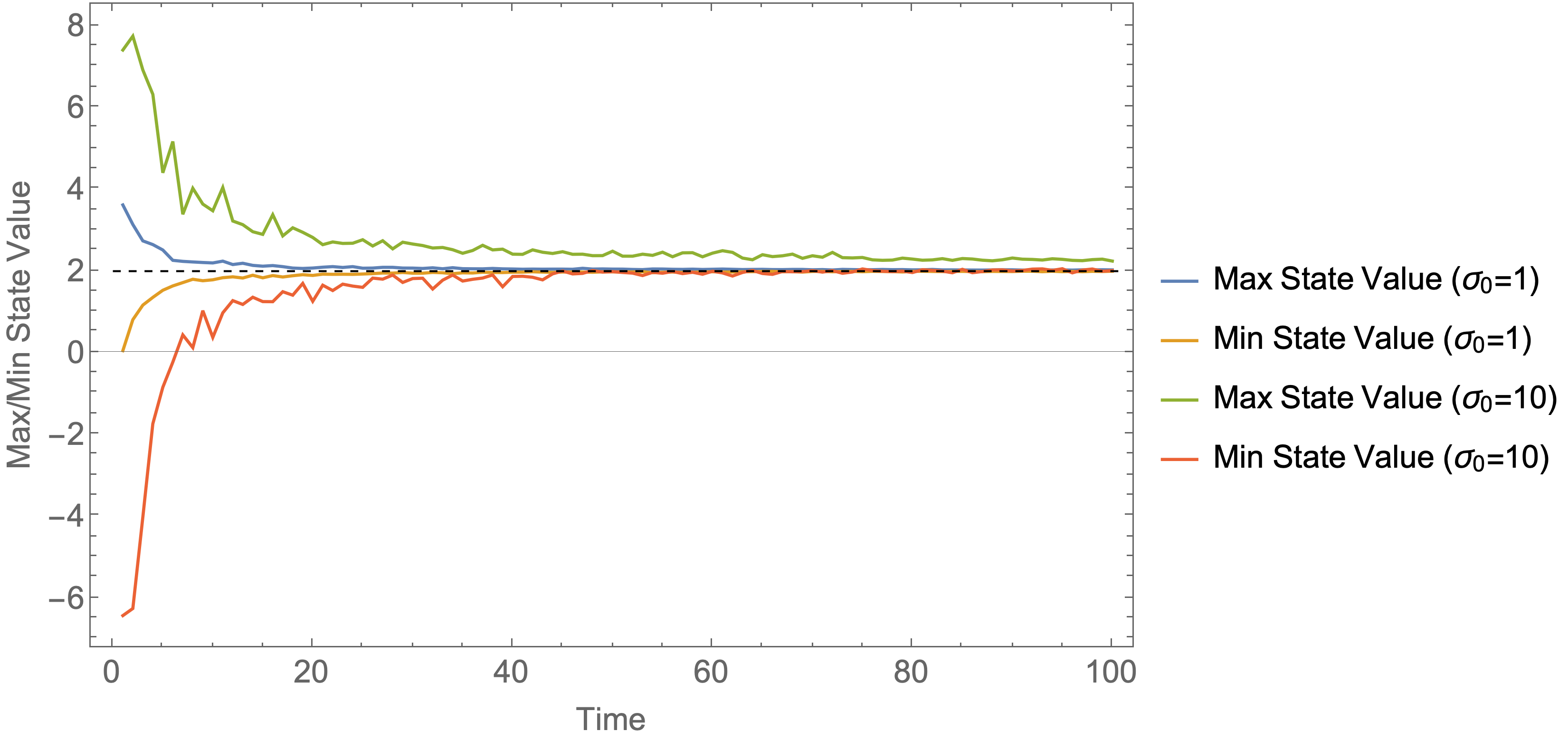}
\caption{Convergence of the system under differing levels of input noise. Larger input noise slows the convergence. As before, only the maximum and minimum state values are shown.}
\label{fig:SlowerConvergence}
\end{figure}

Finally, we illustrate that the rate of convergence is affected (slightly) by $\sigma_0^2$ as expected from \cref{eqn:DerivedUpdate2}. This is shown in \cref{fig:SlowerConvergence}, where we see dynamics generated by larger $\sigma_0^2$ converge more slowly.
This figure was made using a graph with 100 vertices and generated using the Barab\'{a}si-Albert distribution \cite{BA99a} with $2$ edges added in each round. Code for regenerating all results in this section is available in the SI.

These experiments suggest that the hidden state cannot be estimated using data from a single output of the dynamical system but that time series would need to be repeatedly sampled from this system. From a privacy perspective, this implies that if this is used as a consensus mechanism, the hidden state of the agents will be protected statistically. From a scientific perspective, this implies that as users in a social network converge to expressing a common ethos (e.g., on a polarizing topic like climate change or abortion rights) it may be impossible to infer hidden (true) positions among the users. 

\section{Conclusions and Future Work} \label{sec:Conclusions}
In this paper we showed how time-varying peer-pressure can be used to produce consensus in an information exchange system with hidden true states. The consensus value is a weighted average of those hidden states. We also showed that even when the system outputs can be observed and the underlying model is known, the initial state cannot be exactly recovered. This was formally proved in the two agent case and empirically illustrated in a more complex graph. We also showed how the variance of the noise affected the variance of the estimated hidden states, showing it was consistent with the theoretical ansatz.   

In future work it would be interesting to consider a two time-scale dynamic in which the internal (hidden) state is allowed to update at a slower rate as a result of consensus on the public state. Studying cases where the network structure evolve would also be of interest, since the connectivity of the network is a necessary condition for convergence of the system to the consensus value. Additionally, the discrete time dynamic could be replaced with a continuous stochastic differential equation (SDE):
\begin{equation*}
d\mathbf{y} = \left[\mathbf{S}+\rho(t)\mathbf{D}\right]^{-1}\left[\mathbf{S}\left(\mathbf{x}_0 - \mathbf{y}\right) - \rho(t)\mathbf{L}\mathbf{y}\right]dt + \left[\mathbf{S} + \rho(t) \mathbf{D}\right]^{-1}d\bm{\xi},
\end{equation*}
where $\bm{\xi}$ is a vector Wiener process and $L = \mathbf{D} - \mathbf{A}$ is the matrix Laplacian. This SDE is linear and has a well-defined Fokker-Plank equation that could have interesting properties. 

\section*{Acknowledgements}
All authors were supported in part by the National Science Foundation under grant CMMI-1932991. A.S. was supported in part by the National Science Foundation under grant CAREER-1453080.

\appendix

\section{Results on Additional Graphs}
We ran additional experiments using graphs generated with the Barab\'{a}si-Albert distribution \cite{BA99a} with $1$ edge added in each round. Results in this appendix can be compared to those in \cref{sec:Experiment}. The experimental design with this collection of graphs was identical to the experimental design used in \cref{sec:Experiment} except in the density of the graphs investigated. These graphs were half as dense as those studied in \cref{sec:Experiment}.

We fitted the same linear model
\begin{equation}
\langle{\sigma^2}\rangle \sim \beta_0 + \beta_1|V| + \beta_2\sigma_0^2 + \epsilon,
\label{eq:LinModelApp}
\end{equation}
with the new data to obtain the parameter table
\begin{equation*}
\begin{array}{l|llll}
 \text{} & \text{Estimate} & \text{Standard Error} & \text{$t$-Statistic} & \text{$p$-Value} \\
\hline
 1 & -0.92 & 0.19 & -4.77 & 8.56\times 10^{-6} \\
 |V| & 0.00 & 0.00 & 0.37 & 0.71 \\
 \sigma_0^2  & 0.24 & 0.02 & 10.95 & 2.31\times 10^{-17}. \\
\end{array}
\end{equation*}

We can compare this table to the parameter table in \cref{sec:Experiment} by computing the confidence intervals (CI) on each parameter in the two types of experiment. This is shown in \cref{tab:Comparison}.
\begin{table}[htbp]
\begin{tabular}{cc}
\begin{minipage}{0.45\textwidth}
\begin{displaymath}
\begin{array}{l|cc}
 \text{} & \text{Est.}  & \text{CI} \\
\hline
 1 & -0.8  & (-1.1,-0.51) \\
 |V| & 0.00 & (-0.003,0.0017) \\
 \sigma_0^2  & 0.22  & (0.19,0.25) \\
\end{array}
\end{displaymath}
\end{minipage} & 
\begin{minipage}{0.45\textwidth}
\begin{displaymath}
\begin{array}{l|cc}
 \text{} & \text{Est.}  & \text{CI} \\
\hline
 1 & -0.91  & (-1.30,-0.53) \\
 V & 0.000  & (-0.002,0.003) \\
 \sigma_0^2  & 0.24  & (0.2,0.29) \\
\end{array}
\end{displaymath}
\end{minipage}\\
Parameter CI's BA2 Data & Parameter CI BA1 Data
\end{tabular}
\caption{(Left) Parameter confidence intervals for fit of \cref{eq:LinModelApp} on graphs generated with Barab\'{a}si-Albert distribution with 2 edges added per round (BA2). (Right)  Parameter confidence intervals for fit of \cref{eq:LinModelApp} on graphs generated with Barab\'{a}si-Albert distribution with 1 edge added per round (BA1).}
\label{tab:Comparison}
\end{table}

\begin{figure}[htbp]
\centering
\includegraphics[width=0.45\textwidth]{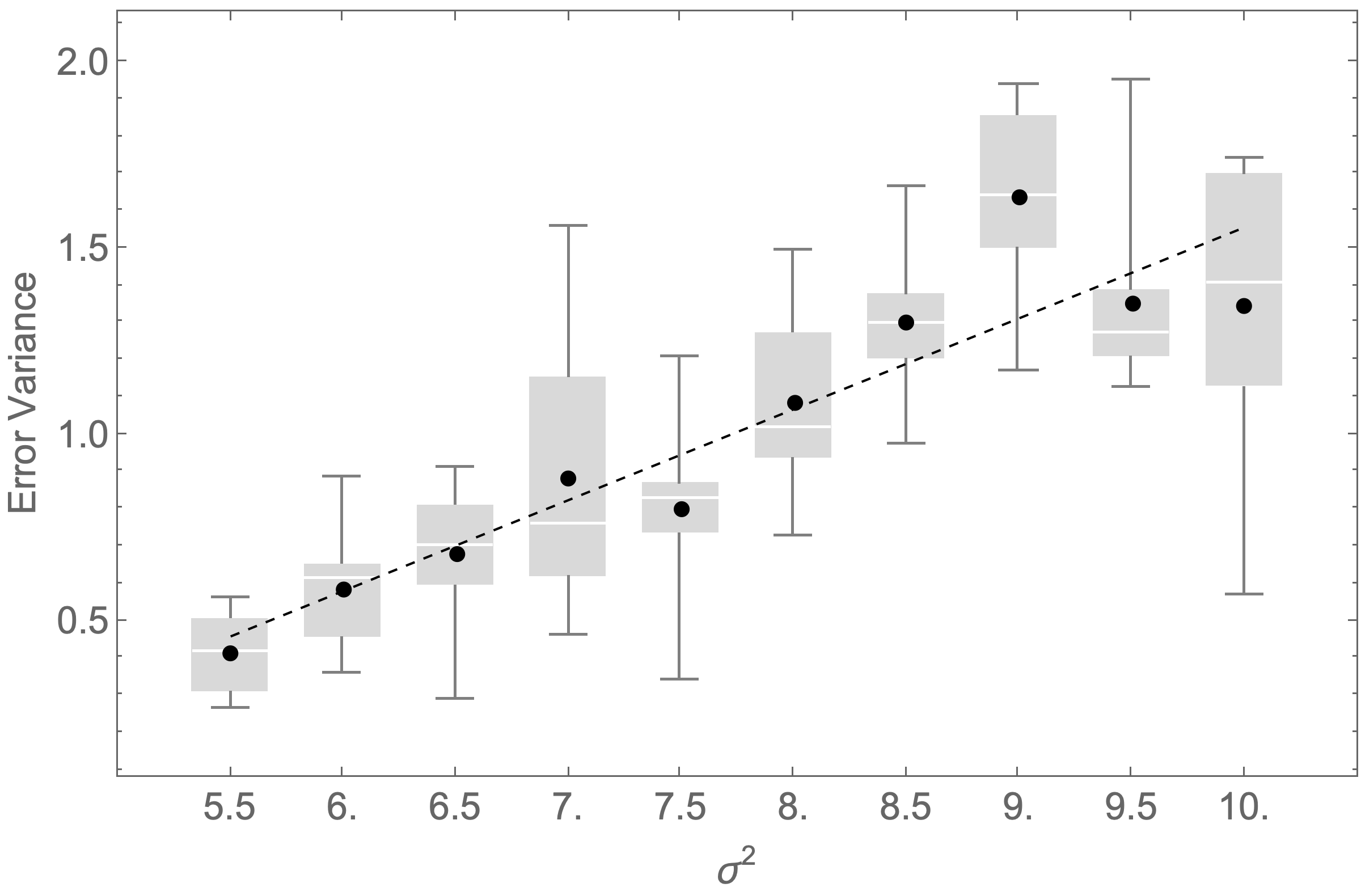}\qquad
\includegraphics[width=0.45\textwidth]{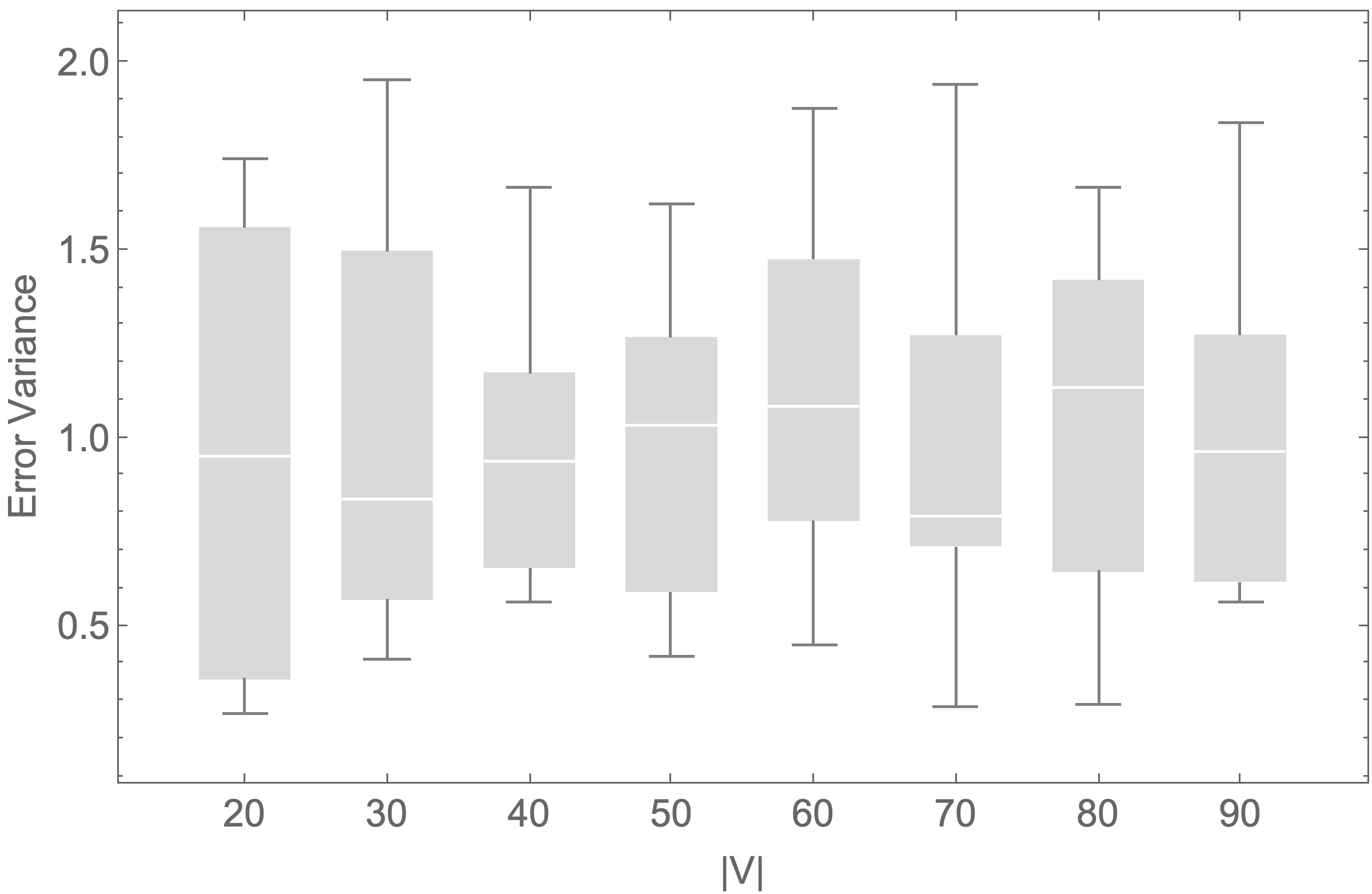}
\caption{(Left) Box-whiskers plot of the residual variances (distribution of $\sigma^2$) over 20 replications for varying noise $\sigma_0^2$ shows a linear relationship between $\hat{\sigma}^2$ and $\sigma_0^2$ as expected. (Right) Box-whiskers plot of the residual variances for varying graph sizes shows no effect from graph size.}
\label{fig:BoxWhiskersBA1}
\end{figure}

\begin{figure}[htbp]
\centering
\includegraphics[width=0.31\textwidth]{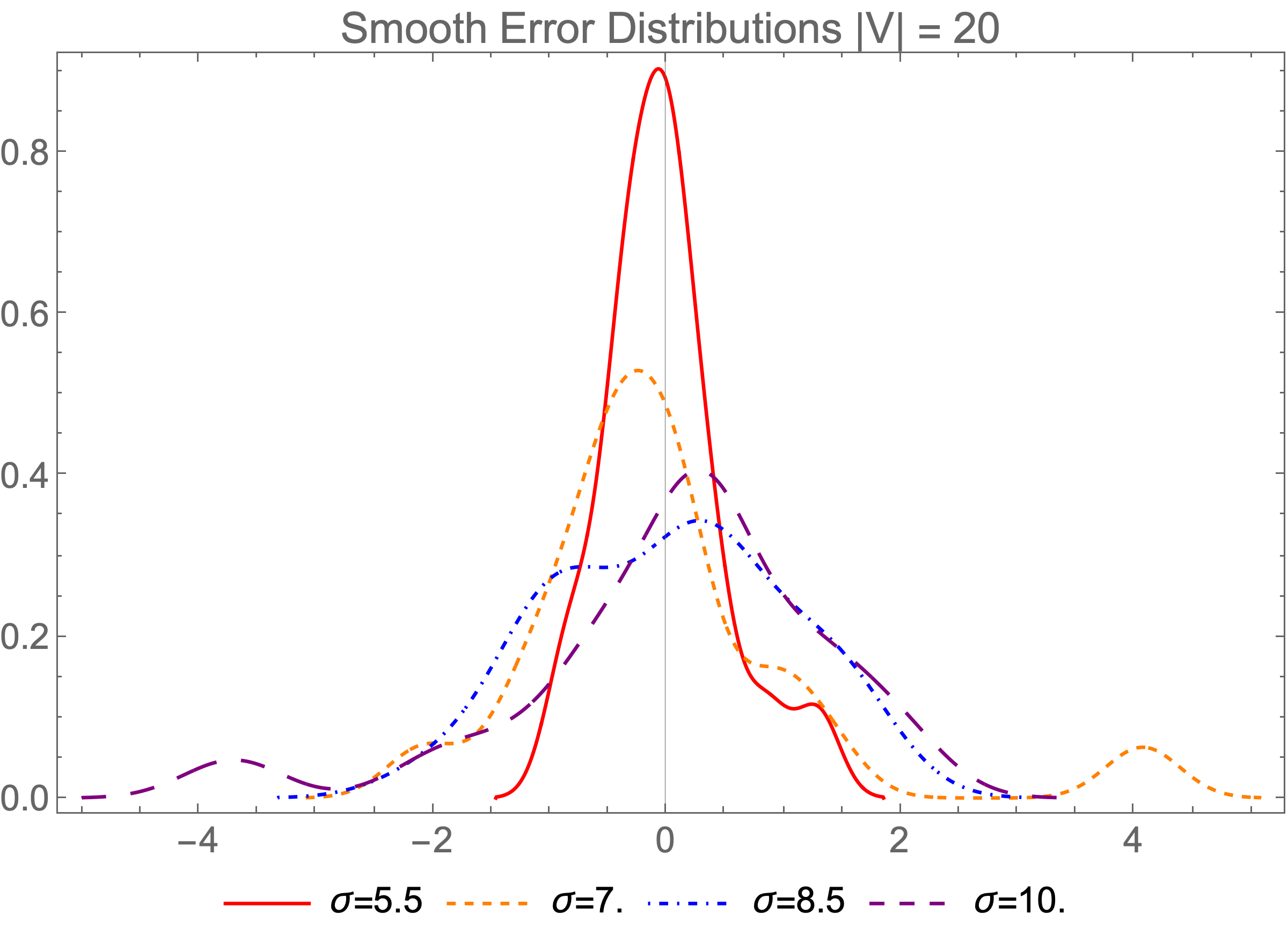}\quad
\includegraphics[width=0.31\textwidth]{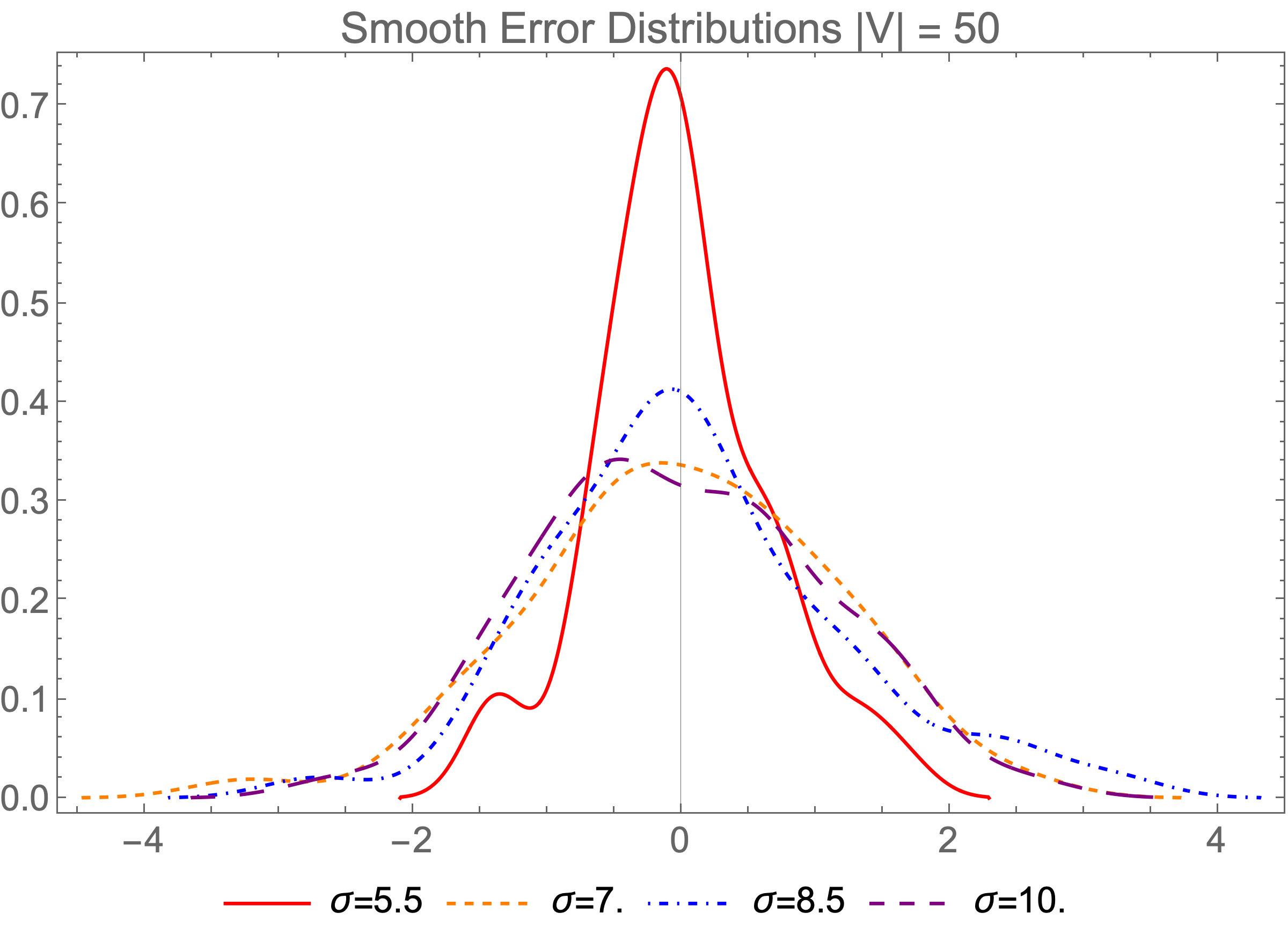}\quad
\includegraphics[width=0.31\textwidth]{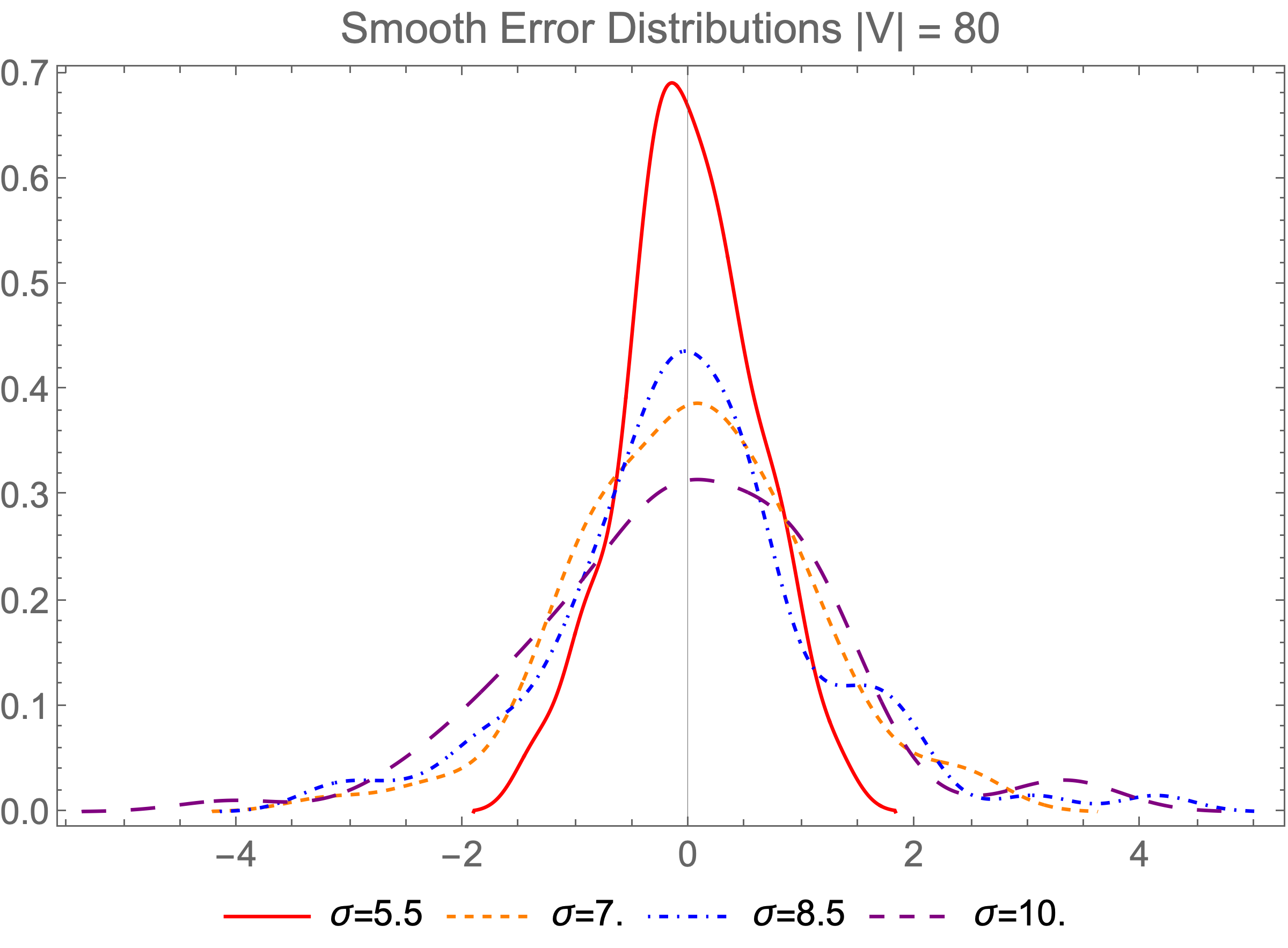}
\caption{(Left) Mean residual distributions for varying $\sigma_0^2$ when $|V| = 20$. (Middle) Mean residual distributions for varying $\sigma_0^2$ when $|V| = 50$. (Right) Mean residual distributions for varying $\sigma_0^2$ when $|V| = 80$.}
\label{fig:Distribs-BA1}
\end{figure}

\cref{tab:Comparison} shows that the parameters in the two graph classes are statistically identical. The data again suggests that the size of the graph has no impact on the residual variance. Again, the mean residual variance $\langle{\sigma^2}\rangle$ grows linearly with noise variance $\sigma_0^2$ as shown in \cref{fig:BoxWhiskersBA1}. This is expected from \cref{eqn:SigmaVariation}.
We also show the effect on mean residual distribution shape as a function of graph size $|V|$ and noise variance $\sigma_0^2$ in \cref{fig:Distribs-BA1} for this set of experiments.
Notice the structures of the distributions are largely similar to each other (unimodal and tending toward normality) with some small-sample effects on the distributions when $|V| = 20$. More importantly, they have the same structure as those presented in \cref{fig:Distribs-BA1}, suggesting that edge density plays a minor roll in the residual error distribution at this scale.

\bibliographystyle{apsrev4-2}
\bibliography{AgentsConcealed}

\end{document}